# Room Temperature Device Performance of Electrodeposited InSb Nanowire Field Effect Transistors

Suprem R. Das[1], Collin J. Delker[2], Dmitri Zakharov[4], Yong P. Chen[1,2], Timothy D. Sands[2,3], and David B. Janes[2,*]

1. *Department of Physics and Birck Nanotechnology Center, Purdue University, West Lafayette, IN 47907, USA*
2. *School of Electrical and Computer Engineering and Birck Nanotechnology Center, Purdue University, West Lafayette, IN 47907, USA*
3. *School of Materials Engineering and Birck Nanotechnology Center, Purdue University, West Lafayette, IN 47907, USA*
4. *Birck Nanotechnology Center, Purdue University, West Lafayette, IN 47907, USA*

## ABSTRACT

In this study, InSb nanowires have been formed by electrodeposition and integrated into NW-FETs. NWs were formed in porous anodic alumina (PAA) templates, with the PAA pore diameter of approximately 100 nm defining the NW diameter. Following annealing at $125^0$C and $420^0$C respectively, the nanowires exhibited the zinc blende crystalline structure of InSb, as confirmed from x-ray diffraction and high resolution transmission electron microscopy. The annealed nanowires were used to fabricate nanowire field effect transistors (NW-FET) each containing a single NW with 500 nm channel length and gating through a 20nm $SiO_2$ layer on a doped Si wafer. Following annealing of the NW-FETs at $300^0$C for 10 minutes in argon ambient, transistor characteristics were observed with an $I_{ON}$ ~ 40 µA (at $V_{DS}$ = 1V in a back-gate configuration), $I_{ON}/I_{OFF}$ ~ 16 - 20 in the linear regime of transistor operation and $g_d$ ~ 71µS. The field effect electron mobility extracted from the transconductance was ~1200 $cm^2$ $V^{-1}$ $s^{-1}$ at room temperature. We report high on-current per nanowire compared with other reported NW-FETs with back-gate geometry and current saturation at low source-drain voltages. The device characteristics are not well described by long-channel MOSFET models, but can qualitatively be understood in terms of velocity saturation effects accounting for enhanced scattering.

* Corresponding author: David B. Janes (janes@ece.purdue.edu)

**INTRODUCTION**

Nanowire field-effect transistors (NW-FETs) have attracted significant research interest in recent years [1-3]. While a number of NW materials have been studied, narrow bandgap materials can provide one-dimensional quantization effects at modest diameter and high mobility, making them suitable candidates for future nanoelectronics applications. Compared to other commonly studied bulk and thin film semiconductors, InSb has a high room temperature carrier mobility, low effective mass, and low (direct) band gap, making it suitable for use in applications such as high-speed, low-power transistors, tunneling field-effect transistors and infrared optoelectronics devices [4-5]. The first reported high-speed and low-power InSb FETs were InSb/In$_{1-x}$Al$_x$Sb enhancement mode devices operating with an $f_{max}$ of 89 GHz at V$_{ds}$ < 0.5V [6]. Recently, Intel and QinetiQ have demonstrated both enhancement and depletion mode InSb quantum well transistors with 305 GHz and 256 GHz unity gain cut-off frequencies, respectively [7]. While NW-FETs operating at comparable frequencies have not yet been demonstrated, the results on planar devices reflect the potential of the material for enabling high-performance devices.

The relatively large lattice mismatch between InSb and commonly available semiconductor substrates presents a significant challenge for development of planar devices employing InSb epilayers. However, NW structures can be grown without a lattice-matched substrate. The Vapor-Liquid-Solid (VLS) growth mechanism, where a metal seed particle of desired diameter dispersed on a substrate acts as a point of initiation of NW growth, has been utilized to grow InSb NWs [8]. InSb NW-FET devices recently reported by Candebat et al and Nilsson et al showed very good performance with ambipolar and *n*-type channel conduction respectively [9-11]. However, repeatable growth of InSb NWs is challenging due to the difficulty in the stoichiometric growth of InSb due to vapor pressure differences of In and Sb and processing temperatures, barrier oxide layer formation around nanowires and various twin defect and stacking fault formation in the structure [12-14]. Despite widespread use of the VLS method, Hannon et al recently demonstrated that in Si NWs there is always a wetting issue arising from the diffusion of Au atoms into the NW sidewalls from the seed particle as the NW growth continues [15]. An alternative growth method that can be very effective and economical uses electrodeposition of the desired ionic species into the nano-channels of a PAA (porous anodic alumina) template with approximately cylindrical pores [16]. The first reported InSb NWFETs using nanowires grown by electrodeposition showed *p*-type channel conduction with a hole mobility of $\mu_h \sim 57$ cm$^2$V$^{-1}$s$^{-1}$ [17]. In this letter, we report NW-FETs employing individual, electrodeposited InSb NWs with high on-current ($I_{ON} \sim 40\mu A$) per NW in comparison to other reported NW-FETs and current saturation at low drain-source voltage ($V_{ds}$). A field-effect electron mobility of $\mu_e \sim 1200$ cm$^2$ V$^{-1}$ s$^{-1}$ was extracted from the room temperature transconductance data.

**EXPERIMENTAL**

Commercial PAA templates (Anodisc 13 from Whatman) with a thickness of 60 μm and a pore diameter that tapered from 20nm to 200nm over the thickness were utilized. A working electrode was formed by e-beam evaporation of 100nm Au on the 20nm pore side of the PAA surface. A Pt-mesh counter electrode and an Ag/AgCl (KCl) reference electrode were employed in a three-electrode electrochemical cell. The InSb nanowires were grown by direct electrodeposition in an aqueous solution containing 0.15M InCl$_3$, 0.1M SbCl$_3$ along with complexion agents 0.36M citric acid and 0.17M potassium citrate, at pH 1.8 at a reduction potential of -1.5V [16]. Using different deposition times (20 minutes to 1 hr), several samples of different nanowires length were grown. The as-deposited nanowires were annealed in argon atmosphere at 125$^0$C for 6hr and then at 420$^0$C for 4hr. [16]. Following annealing, the PAA template was dissolved in 1M KOH solution (overnight). The nanowires attached to the gold layer were then released into isopropyl alcohol (IPA) via sonication. The electron microscope images of the nanowires were captured using a Hitachi Field Emission Scanning Electron Microscope (FESEM) and FEI Titan 80-300kV Transmission Electron Microscope (TEM).

Back-gated NW-FETs were fabricated using a silicon substrate as the gate contact and a thermal oxide as the gate dielectric, as shown in Figure 2. A heavily doped (p+) silicon wafer with 20nm SiO$_2$ was patterned with an array of alignment markers using e-beam lithography, e-beam evaporation, and liftoff of nickel. The nanowires were then dispersed onto the wafer by casting a few drops of the NW suspension onto the wafer surface and allowing the solvent to evaporate. Wires were located in reference to the alignment markers via SEM imaging. Source and drain contacts were then individually aligned to each wire and defined by an e-beam lithography, nickel deposition (90 nm), and liftoff process. Channel lengths ($L_{CH}$) and contact lengths were both 500nm. The current-voltage (I-V) characteristics of the devices were measured using a semiconductor parameter analyzer (Model HP4155C).

**RESULTS AND DISCUSSIONS**

Figure 1 shows a High Resolution TEM (HRTEM) image of one of the electrodeposited InSb NWs. The wires were found to be single crystalline without any visible dislocations, twin defects or stacking faults. The magnified image shows the lattice planes of the NW along with the [111] growth direction. The inset shows the selective area electron diffraction (SAED) pattern obtained from FFT analysis, which indicates a crystal structure consistent with that of bulk InSb without any crystal defects. X-ray diffraction indicated a zinc blende structure with a lattice constant of 6.4782 A$^0$. The wires were generally found to be tapered, consistent with the tapering in the PAA template. Near the narrow end, a diameter of ~100 nm is obtained over a length of ~2 μm. The NW-FET fabrication process allowed selection of specific regions along the length of the nanowires, so the channel regions of the NW-FETs were defined in regions of the NW with diameters of 90 - 100nm. Fig. 2 shows the FESEM image of a representative NWFET device with gate length ($L_g$) of 500 nm set by the S/D spacing and typical Ω-type 500 nm wide Ni S/D contact electrodes. The actual device (on which the present data is shown) has a uniform diameter of around 100 nm as confirmed from the FESEM image.

The measured n-channel, room temperature output characteristics of the device are shown in Fig. 3. Fig. 3(a) shows the output characteristics of the as-fabricated device and 3(b) shows the characteristics after the devices are annealed at 300$^0$C in argon ambient for 10 minutes. Before annealing the representative devices showed no visible current saturation, channel resistance ($R_{ds}$) of 90kΩ for small $V_{ds}$ and a gate-source voltage ($V_{gs}$) = 2V, and heavily contact-dominated output characteristics (indicated by the diode-like characteristics). After annealing, the low field $R_{ds}$ at $V_{gs}$ = 2V improved to 14kΩ (indicating reduction of contact effects upon annealing) and the devices exhibited I-V characteristics more consistent with channel-dominated behavior including saturation. Similar effects of decreasing the contact resistance upon annealing were reported earlier [18]. The on-currents were found to be around 40μA for drain voltages of 1V, significantly higher than reported values on wider bandgap materials. The extracted transconductance data in the linear regime of operation is shown in Fig. 4 (a) and the peak $g_m$ of 2.75μS for $V_{ds}$ = 100mV was obtained from the plot. Using a cylinder-on-plate capacitance model for the gate (since $d_{NW} \gg t_{ox}$) and the peak transconductance value [18], an effective electron field-effect mobility of $\mu_e$ ~ 1200 cm$^2$V$^{-1}$s$^{-1}$ was extracted. This value is significantly lower than reported values for bulk InSb, indicating that contact effects or scattering mechanisms such as boundary scattering play a significant role. In the saturation region ($V_{ds}$ = 1V), a peak transconductance of 16μS was extracted. The output resistance in saturation was ~ 75 kΩ, which leads to an intrinsic gain (Gm x Rds) of 1.2

In a long-channel MOSFET model, $I_d$ saturation corresponds to channel pinch-off at the drain end of the channel. In this case, the saturation (knee) voltage is $V_{knee} = (V_{gs} - V_{th})$. In our devices, current saturation occurred at $V_{ds}$ values around 500 mV even for ($V_{gs} - V_{th}$) ~ 3V, as compared to 100 mV saturation reported by Nilsson et al [10]. This early onset of saturation indicates that a mechanism other than channel pinch-off is responsible for the saturation. We did not observe hard saturation in the drain current up to $V_{ds}$ = 1V. The increase in $I_d$ with increasing $V_{ds}$ (beyond $V_{knee}$) can be explained, at least in part, by modulation of the channel potential (particularly near the source) by the drain. Due to 1-D electrostatics for the case in which the gate overlaps the S/D regions, the band bending near the source and drain is exponential with a characteristic length of $\lambda = \sqrt{d_{wire} d_{ox} \varepsilon_{wire}/\varepsilon_{ox}} \approx$ 90nm [20]. Although a channel length of ~ 5λ should be long enough to avoid substantial short channel electrostatic effects, a modest modulation of the source barrier is expected. In addition, fringing fields from the extended drain contact will modulate the channel potential. As the ratio between the channel length and the body (NW) thickness is increased (e.g. by decreasing the NW diameter), the drain modulation would be expected to decrease.

Fig. 4 (a) illustrates the transfer characteristics of the annealed devices (annealed at $300^0$C in argon for 10 minutes) in the linear regime at drain voltages of 50mV and 100mV respectively with an $I_{ON}/I_{OFF}$ ~ 16 – 20. These measurements were done following the high $V_{ds}$ measurements, which may have caused a bias-induced stress in the devices that shifted the threshold voltage to a more negative value. Subthreshold swings of several volts per decade were observed, which can be explained in part by the relatively thick body (NW diameter) but also likely indicating the presence of a high level of interface traps between the oxide and nanowires which is common for this bottom-gate geometry, particularly on $SiO_2$.

Figure 4(b) shows the transfer characteristics of the annealed devices in the saturation regime at drain voltages of 0.5V and 1V respectively. An $I_{ON}/I_{OFF}$ ratio of ~ 12 - 15 was observed and the threshold voltage was estimated near -1V. Figure 4 (b) inset shows the expected energy band diagram along the 500nm channel at source-drain voltage of 0.5V and both at low and high gate bias points, reflecting the exponential potential profiles described earlier. Due to the low bandgap, ambipolar behavior starts to appear at this drain voltage and higher ($V_{ds}$ of 1V also shown) and gate voltages below threshold. In this bias regime, a thin drain barrier allows electrons to tunnel out of the valence band into the drain, increasing the current. This is consistent with the measured transfer characteristics which show slight increase in current at high $V_{ds}$ and low $V_{gs}$, but no drastic increase at lower $V_{ds}$ (Figure 4a). The bandgap of the wire can be approximated from the $I_{ON}/I_{OFF}$ ratio, using the relationhip $(I_{ON}/I_{OFF}) = \frac{1}{2} \exp(E_g/2kT)$. While this relationship is strictly correct for a one-dimensional channel, ballistic transport, and low drain voltages, it can provide an estimate of bandgap for devices in which both electrons and holes can be readily injected and heavily doped "body" regions are not present. Using this approximation, the bandgap was estimated to be ~ 0.20eV, which is consistent with the accepted value of 0.175eV for crystalline InSb.

As mentioned previously, a mechanism other than channel pinch-off is required to explain the early onset of saturation. The carrier velocity can be estimated and compared to the bulk saturation velocity for InSb. The number of carriers in the channel per unit length at $V_{gs} = 2V$ can be estimated by the gate capacitance (cylinder-on-plate) and gate voltage to be ~ $4.7 \times 10^7$ cm$^{-1}$. From this number, in conjunction with the measured saturation current of 40μA, an average electron velocity can be calculated to be ~ $5.3 \times 10^6$ cm/s. This is significantly lower than the $4 \times 10^7$ cm/s saturated drift velocity of bulk InSb [21]. While detailed calculations of saturated velocities in nanowires are not available, a reduced saturated velocity could result from enhanced scattering (surface, electron-phonon and electron-electron) in the nanowires and is generally consistent with the observation of mobility lower than reported bulk values. For velocity saturation, one would generally expect the critical electric field ($E_{crit}$) to be related to the saturated velocity ($v_{sat}$) by $v_{sat} \sim \mu E_{crit}$. Assuming that our inferred velocity corresponds to $v_{sat}$, and using our measured low-field mobility, we estimate $E_{crit}$ ~ 5,000 V/cm. The electric field at the onset of saturation ($V_{knee} / L_{CH}$ ~ 10 kV/cm) is consistent with this estimate, indicating that a velocity saturation model with reduced $v_{sat}$ may be valid.

While InSb nanowires are good candidates for both operation in a single 1-D sub-band and ballistic operation, the NW-FETs in the current study are not expected to exhibit significant effects from these mechanisms. The spacing between the first two sub-bands is approximately equal to the thermal energy (kT) at room-temperature. Coupled with operation at $V_{ds}$ values larger than kT/q, a number of sub-bands should be partially populated and participate in conduction. The room temperature ballistic mean free path of the electrons in bulk InSb is reported to be 580 nm [6]. Based on the channel length in the present study (500 nm), one might expect the devices to be operating in a semi-ballistic transport regime. However, the measured transconductance per conductive mode in our NW-FET is significantly smaller than the fundamental conductance quantum $G_0$ ($G_0 = 2e^2/h = 77.5$ μS), indicating that the devices are not in the ballistic limit. This is consistent with the observation of mobility significantly lower than the bulk value, and the associated interpretation in terms of enhanced scattering mechanisms.

## CONCLUSIONS

In summary, n-channel InSb NWFETs were fabricated using single crystalline electrodeposited InSb NWs. The measured room temperature I-V characteristics show low on-resistance, high drain current and saturation at low $V_{ds}$. An $I_{ON}/I_{OFF}$ ratio of ~ 20 in the linear and saturation regimes and an electron mobility of 1200 cm$^2$V$^{-1}$s$^{-1}$ were achieved. The $I_{ON}/I_{OFF}$ ratio and ambipolar behavior are both indicative of the low bandgap of the InSb nanowires. Further improvements in the device metrics can be achieved by creating better ohmic contacts, reducing interface traps in the gate-oxide interface, and improving the gate geometry with a top-gate or wrap-around gate structure.

## ACKNOWLEDGEMENTS

We acknowledge Y. Zhao, D. Candebat, Prof. Chen Yang and Prof. Joerg Appenzeller for helpful discussions. This work was supported in part by the Midwest Institute for Nanoelectronic Discovery (MIND) and by DARPA.

**FIGURE CAPTIONS**

Fig. 1 High Resolution Transmission Electron Microscope (HRTEM) image of an InSb nanowires and magnified view showing the individual lattice planes. Inset shows the SAED pattern of the InSb NW

Fig. 2 Field Emission Scanning Electron Microscope (FE-SEM) image showing a representative InSb NWFET device with Ni S/D and back gate structure.

Fig. 3 Output characteristics ($I_d$-$V_{gs}$) of (a) the as-fabricated device and (b) device after annealing in argon

Fig. 4 (a) Transfer characteristics ($I_d$-$V_{gs}$) and the transconductance versus gate voltage of the annealed device in linear regime ($V_{ds}$ of 50mV and 100mV). (b) Transfer characteristics ($I_d$-$V_{gs}$) of the annealed device in saturation regime. Fig. 4(b) inset illustrates the band diagram of 500nm channel InSb NWFET at $V_{ds}$ = 0.5V with sub-threshold and above threshold gate voltages.

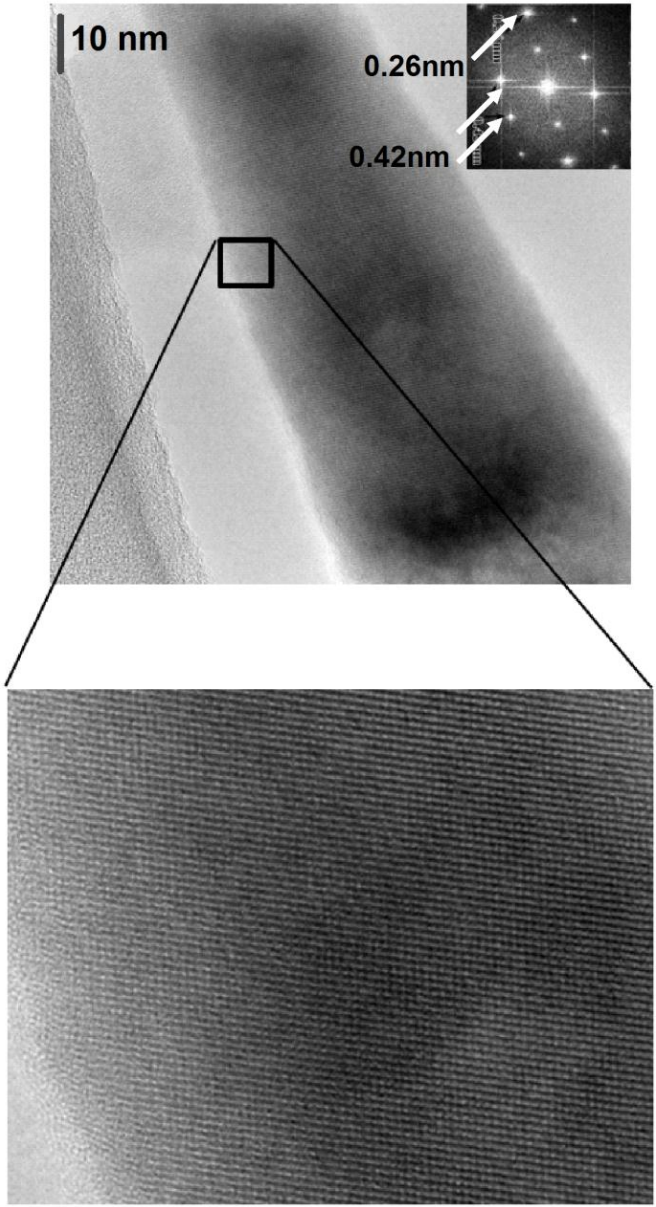

**Figure 1: Das et al**

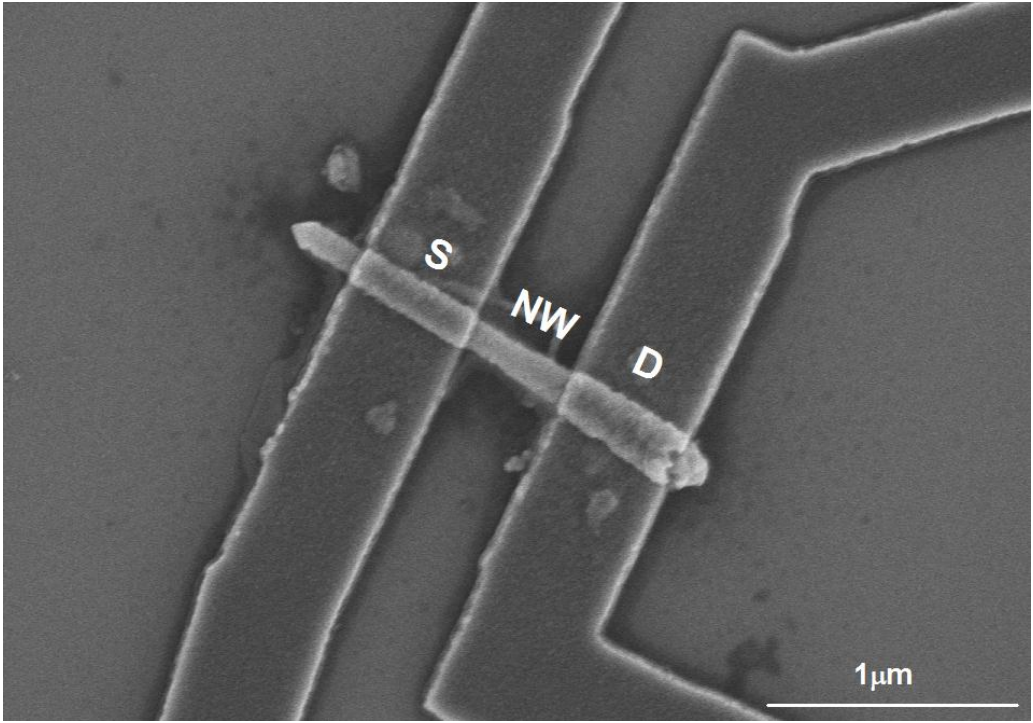

**Figure 2: Das et al**

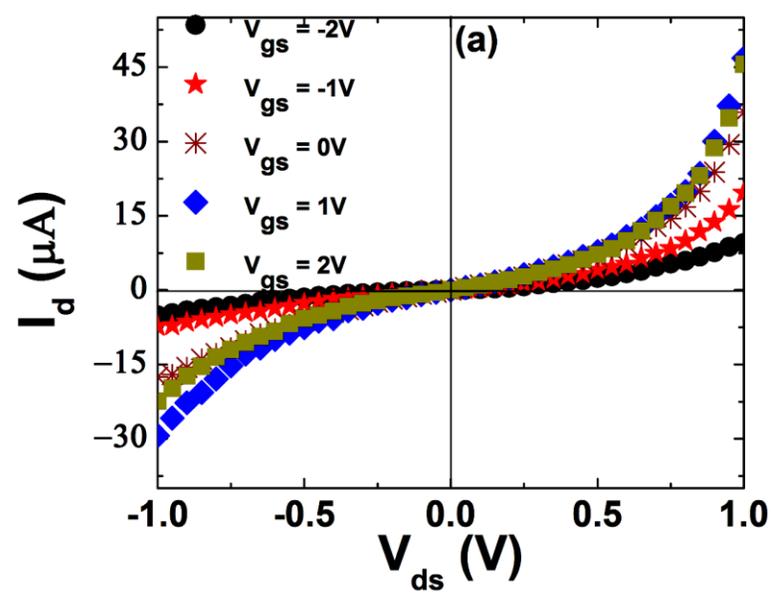

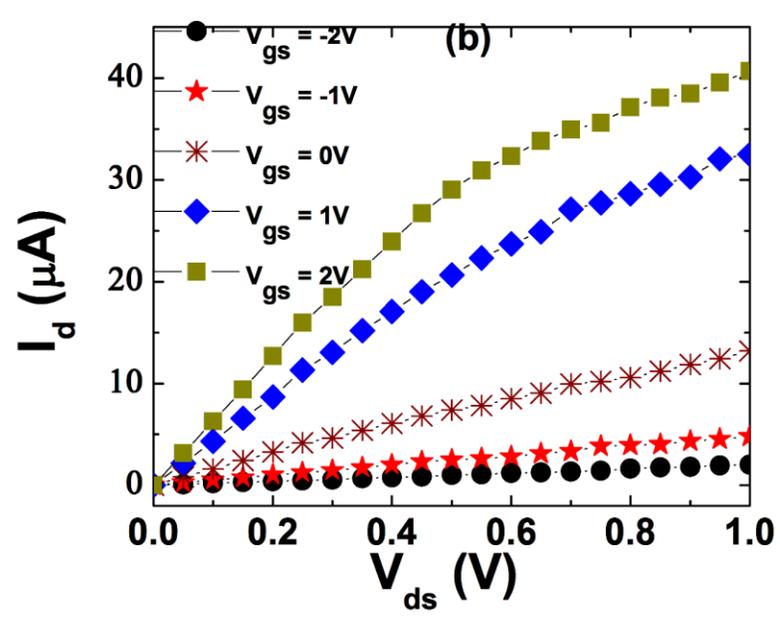

**Figure 3: Das et al**

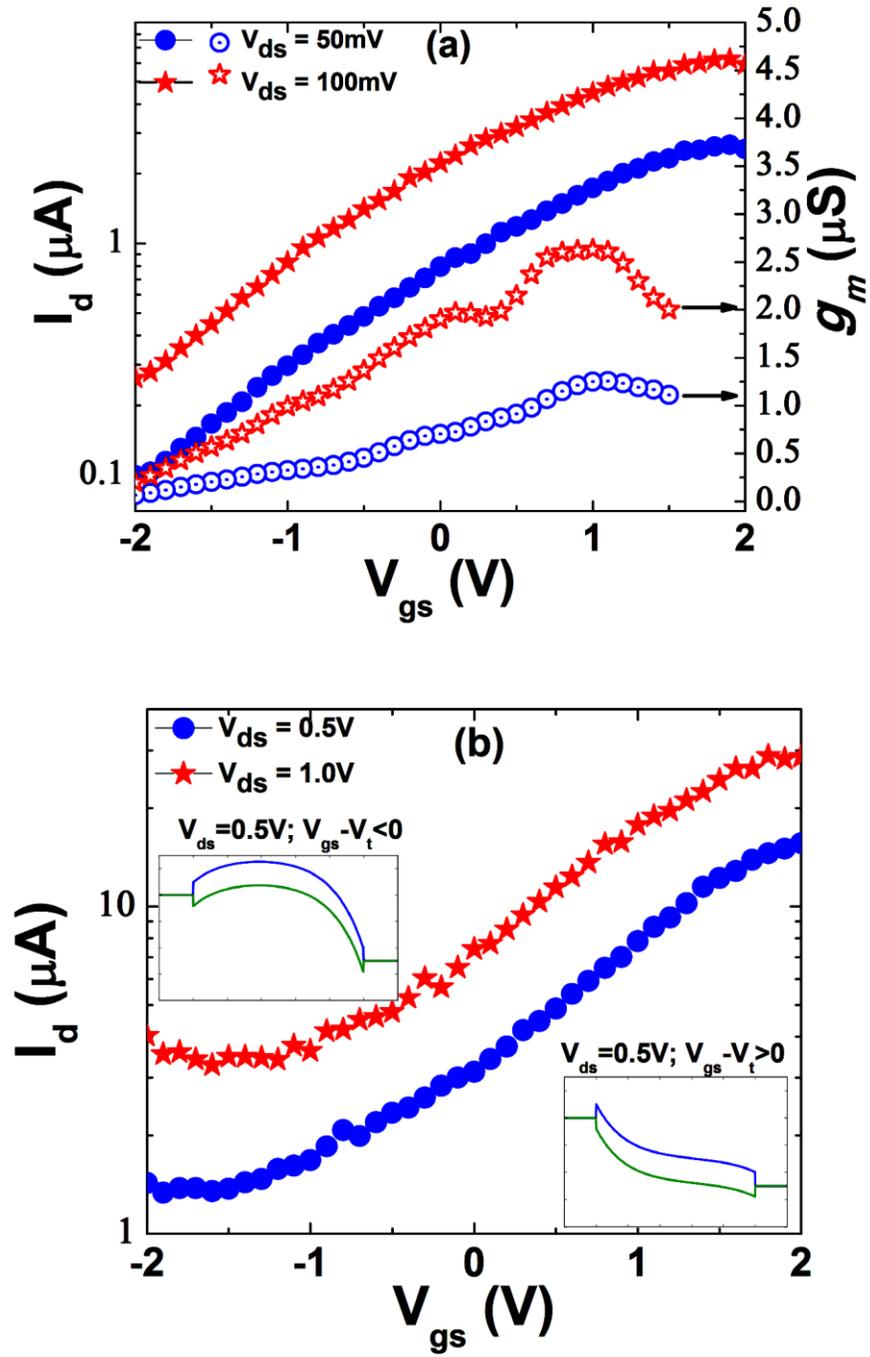

**Figure 4: Das et al**